\documentstyle[epsf]{article}
\title{Nonhomogeneous magnetization and superconductivity in
  superconductor-ferromagnet structures}
\author{F.S.Bergeret $^{1 }$\thanks{Corresponding Author: F.S.Bergeret; Theoretische Physik III,
Ruhr-Universit\"{a}t Bochum, D-44780 Bochum, Germany; tel:+49-234-3225132;
fax: +49-234-32214448; e-mail: sebas@tp3.ruhr-uni-bochum.de.} , A.F. Volkov$^{1,2}$ and \underline{K.B.Efetov}$^{1,3}$\\$^{(1)}$Theoretische Physik III,\\
Ruhr-Universit\"{a}t Bochum, D-44780 Bochum, Germany\\
$^{(2)}$Institute of Radioengineering and Electronics of the Russian Academy%
\\
of Sciences, 103907 Moscow, Russia \\
$^{(3)}$L.D. Landau Institute for Theoretical Physics, 117940 Moscow, Russia\\
 }
\date{}

\begin{document}
\maketitle

\begin{abstract}
We study two different superconductor-ferromagnet (S/F) structures. We
consider first a Josephson junction which consists of two S/F bilayers
separated by an insulating layer. We show that for an antiparallel alignment of
the  magnetization in the two F layers the Josephson critical current
$I_c$ increases with increasing exchange field $h$. The second system we
consider is a S/F structure with a local inhomogeneity of the magnetization in
the ferromagnet near the S/F interface. Due to the proximity effect not only a
singlet but also a triplet component of the superconducting condensate is
induced in the ferromagnet. The latter penetrates over the length
$\sqrt{D/\epsilon}$ ($D$ is the diffusion coefficient and $\epsilon$ the
energy).  In the case of temperatures of the order of the Thouless energy this
length is comparable to the length of the ferromagnet. This long-range
penetration leads to a significant increase of the ferromagnet conductance
below the superconducting critical temperature $T_c$. Contrary to the case of the 
singlet component, the contribution to the conductance due to the odd 
triplet component is not zero at $T = 0$ and $V = 0$ ($V$ is the voltage) and 
decays with increasing temperature T in a monotonic way.
\end{abstract}



\section{Introduction}

The interplay between superconductivity and ferromagnetism has been the
subject of  extensive research for many years. The ferromagnetism, being
usually much stronger than superconductivity, is supposed to destroy the
latter. This suppression is caused by two mechanisms. One of them is related
to the internal magnetic field which is created by the ordered magnetic
moments. The internal magnetic field, which is proportional to the
magnetization $M$, induces Meissner currents and suppresses the
superconducting order parameter $\Delta$. This mechanism was first analyzed
by Ginsburg \cite{Gins}. He concluded that the coexistence phase could not
take place in an ordinary bulk sample. The second mechanism is due to the
direct action of an exchange field $h$ on spins of electrons\cite{Matthias}.
The exchange field tends to align the spins of the electrons in one
direction and thereby destroys the singlet Cooper pairs. The influence of the
magnetization on the orbital condensate motion may be reduced drastically if
the magnetization changes its direction on the scale of the superconducting
correlation length in such a way that the averaged magnetic field is zero
(as in the case of a spiral magnetic structure). The action of the exchange
field may be also decreased if it is realized via the RKKY interaction \cite{AndSuhl} ( for details see the review \cite{Bul}). The coexistence of
superconductivity and ferromagnetism was observed in ternary rare earth
compounds \cite{Moncton}.

The layered superconductor/ferromagnet (S/F) structures open new
possibilities to achieve the coexistence of ferromagnetism and
superconductivity. In such systems ferromagnetic and superconducting regions
are spatially separated; therefore if the magnetization is oriented parallel
to the layers, it does not strongly affect  the condensate in the S layers.
On the other hand the Cooper pairs penetrates into the F layer over a length
which  in the dirty limit ( $h\tau\ll1$, $\tau $ is the elastic scattering
time) is of the order of $\sqrt{D/h}$, where $D$ is the diffusion
coefficient in the F layer. The S/F structure becomes superconducting if the
F layers are thin enough. Layered structures can be created artificially or
can be found in some high-$T_c$ cuprates, as in the compound $%
RSr_{2}RCu_{3}O_{8}$ ($R$: rare earth). Its structure is similar to the one
of the high-$T_c$material $YBa_{2}Cu_{3}O_{7}$. Magnetic ordering in $RuO_{2}
$ layers (it is assumed that mostly an antiferromagnetic order is realized)
occurs at $T_M=133K$ and superconducting transition presumably in $CuO_{2}$
layers occurs at $T_{c}\simeq 50K$ \cite{cuprate1}. It is interesting to
note that another material from the rutinate class, $Sr_{2}RuO_{4},$ is an
exotic triplet superconductor \cite{Maeno}. A few interesting effects were
predicted and observed in layered S/F structures. Since the experimental
pioneering work performed by Werthamer et al. \cite{Wert}, it is known that
the superconducting critical temperature $T_{c}$ is lowered in S/F
multilayered structure with increasing thickness of the F layers. This is
due to the proximity effect, i.e. superconductivity in the S layer is suppressed
to some extent if it is brought into contact with a non-superconducting
(specially magnetic) layer. Another interesting phenomenon which can occur
in S/F layered structures is an oscillatory dependence of the critical
temperature $T_{c}$ and the critical current on the thickness of the F layer 
\cite{Bul1,Clem,Buzdin1,Jiang}. These oscillations are related to
oscillations of the condensate in the F layers which can be formally
obtained by the replacement $\omega \rightarrow $ $\omega +ih$ (here $\omega 
$ is the Matsubara frequency which enters in the expression for the
condensate Green's functions). An interesting behavior of the critical
Josephson current $I_{c}$ as a function of the exchange field $h$ was
predicted first by Bulaevskii et al. \cite{Bul1}; they found that with
increasing $h$ the current $I_{c}$ changes sign and the phase difference $%
\varphi =\pi $ is established in a Josephson S/F/S junction (the so called $%
\pi $-junction). This prediction was confirmed in a recent experiment \cite
{Ryaz}.

Despite of a large number of theoretical works on equilibrium and
nonequilibrium effects in S/F/S structures, some important features of the
effects remain unclear. In this report we discuss the influence of different
relative orientations of the magnetizations on the critical  Josephson current in a
tunnel SF/I/FS\ junction. Its electrodes consist of SF bilayers. The
critical current $I_{c}$ is not zero if the thickness of the ferromagnetic F
layer is thin enough and depends on relative orientation and absolute values
of magnetization or exchange energy $h$. Surprisingly the current $I_{c}$
turns out to be not decreasing, but increasing function of $h$ in the case
of antiferromagnetic configuration (the magnetization vectors are aligned in
opposite directions in the electrodes). Therefore the critical current at
nonzero $h$ exceeds its value in the absence of $h$.

The second problem which we discuss here is a possible mechanism of the
enhanced conductance measured experimentally in mesoscopic S/F structures.
Recent experiments on  S/F structures showed that below the critical
temperature $T_{c}$ the conductance of the ferromagnetic wire (or film)
varies with the temperature in a nonmonotonic way and may exceed its value above 
$T_{c}$ \cite{Petr,Pann,Chandr}. The decrease of the conductance, which was also observed, has been explained in a few theoretical papers \cite{Been,Falko,Naz}. Although it was assumed in some papers \cite{Been,Naz} that an increase in the conductance may be due to
scattering at the S/F interface,  careful measurements demonstrated that
the entire change of the conductance is due to an increase of the
conductivity of the ferromagnet \cite{Petr,Pann}. Such an increase would not
be a great surprise if instead of the ferromagnet one had a normal metal N.
It is well known (see for review \cite{Been,Lamb}) that in S/N structures the
proximity effect can lead to a considerable increase of the conductance of
the N wire provided its length does not exceed the phase breaking length $%
L_{\varphi }$. However, in a S/F structure if the superconducting pairing is
singlet, the proximity effect is negligible at distances exceeding a much
shorter length $\xi _{h}\sim \sqrt{D/h},$ where $h$ is the exchange energy
in the ferromagnet. This reduction of the proximity effect due to the
exchange field of the ferromagnet is clear from the picture of Cooper pairs
consisting of electrons with opposite spins. The proximity effect is not
considerably affected by the exchange energy only if the latter is small
enough, {\it i.e.} $h<T_{c}$. In the experiments of Refs.\cite{Petr,Pann}
strong ferromagnets, as  $Ni$ or $Co$, were used. Their exchange energy $h$ is
by several orders of magnitude larger than $T_{c}$, and therefore a singlet pairing is
impossible due to the strong difference in the energy dispersions of the
two spin bands.  At the same time, an arbitrary exchange field cannot destroy
a triplet superconducting pairing because the spins of the electrons forming
Cooper pairs are already parallel. We suggest a new mechanism for the
increase of the conductance in S/F structures. This mechanism is based on
the formation of the triplet component in the F wire, which is due to a
local inhomogeneity of the magnetization $M$ in the vicinity of the S/F
interface. We show that the inhomogeneity generates a triplet component of
the superconducting order parameter with an amplitude comparable with that
of the singlet pairing. The penetration length of the triplet component into
the ferromagnet is equal to $\xi _{\varepsilon }=\sqrt{D/\varepsilon }$,
where the energy $\varepsilon $ is of the order of temperature $T$ or the
Thouless energy $E_{T}=D/L^{2},$ $L$ is the sample size. The length $\xi
_{\varepsilon }$ is of the same order as that for the penetration of the
superconducting pairs into a normal metal and therefore the increase of the
conductance due to the proximity effect can be compared to that in an
S/N structure. The inhomogeneity may appear in a natural way as an edge
effect or may be created intentionally. The triplet component we considered differs from that in $Sr_{2}RuO_{4}$; it is almost symmetrical in
momentum $p$ (the antisymmetrical part is small) and is an odd function of
the Matsubara frequency $\omega $. This type of the odd triplet component
was  for the first time suggested by Berezinskii \cite{Berez} as a possible
condensate phase in superfluid He$^{3}$ (in fact a condensate function
symmetrical in $\omega $ and antisymmetrical in momentum $\overrightarrow{p}$
takes place in He$^{3}$). Later the so called odd superconductivity was
discussed in Refs.\cite{Balat} as a possible mechanism for high $T_{c}$
superconductivity. This odd in $\omega $ and even in $\overrightarrow{p}$
condensate component is not suppressed by the impurity scattering.

\section{The SF/I/FS junction}

In this section we consider a layered system consisting of two F/S bilayers
separated by an insulating layer (see Fig.1). In this case the Josephson
critical current is determined by the transparency of the insulating layer
and depends on the relative orientation of magnetization in the F layers.
We assume that the F and the S layers $d_{F,S}$ are thin enough: $%
d_{F,S}<\xi _{F,S}$, where $\xi _{F}=\sqrt{D/h}$ and $\xi _{S}=\sqrt{%
D/\Delta }$. With this assumption  one comes to effective values of the superconducting
order parameter $\Delta _{eff}$, of the coupling constant $\lambda _{eff}$,
and of the magnetic moment $h_{F}$ described by the following equations 
\[
\Delta _{eff}/\Delta =\lambda _{eff}/\lambda =\nu _{s}d_{s}\left( \nu
_{s}d_{s}+\nu _{f}d_{f}\right) ^{-1}, 
\]
\begin{equation}
h_{F}/h=\nu _{f}d_{f}\left( \nu _{s}d_{s}+\nu _{f}d_{f}\right) ^{-1}
\label{a1}
\end{equation}
where $\nu _{s}$ and $\nu _{f}$ are the densities of states in the
superconductor and ferromagnet, respectively.

 First, we analyze the case of a high S/F interface
transparency, i.e. $R_{S/F}<\rho _{F}/\xi _{F}$. Under these conditions all
the Green's functions are nearly constant in space and continuous across the
S/F interface.

In order to find the Green's functions we use the Usadel equation
 \cite{Usadel}, which in the presence of a nonhomogeneous exchange field has
 the general form

\begin{equation}
-iD\nabla \left( \check{{\bf g}}\nabla \check{{\bf g}}\right) +i\left( \hat{
\tau}_{3}\otimes \hat{\sigma}_{0}.\partial _{t}\check{{\bf g}}+\partial
_{t^{\prime }}\check{g}.\hat{\tau}_{3}\otimes \hat{\sigma}_{0}\right) +eV(t) 
\check{{\bf g}}-\check{{\bf g}}eV(t^{\prime })+\left[ \hat{\Delta}\otimes 
\hat{\sigma}_{3},\check{{\bf g}}\right] + \left[\check{M}_{h},\check{{\bf g}}%
\right]\!\! =\!\!0   \label{usadel}
\end{equation}

Here $\check{{\bf g}}$ is the quasiclassical Green's function, which has the
form
\begin{equation}
\check{{\bf g}}=\left( 
\begin{array}{cc}
\check{g}^{R} & \check{g}^{K} \\ 
0 & \check{g}^{A}
\end{array}
\right) \; ,  \label{matrix}
\end{equation}

$\check{M}_{h}=h\left(\hat{\tau}%
_{3}\otimes \hat{\sigma}_{3}\cos \alpha +\hat{\tau}_{0}\otimes \hat{\sigma}%
_{2}\sin \alpha \right)$,

\[
\hat{\Delta}=\left( 
\begin{array}{cc}
0 & \Delta  \\ 
-\Delta ^{\ast } & 0
\end{array}
\right) \;,
\]
 and the matrices $\hat{\tau}_{i}$ and $\hat{\sigma}_{i}$ are the Pauli matrices
in the Nambu and spin space respectively; $i=0,1,2,3$, where $\hat{\tau}_{0}$
and $\sigma _{0}$ are the corresponding unit matrices. The Eq.(\ref{usadel}) is supplemented by the normalization condition 
\begin{equation}
\check{{\bf g}}.\check{{\bf g}}=\check{1}\;.  \label{normalization}
\end{equation}

The current density is determined by the
usual expression 
\begin{equation}
I_{J}=\frac{1}{16\rho }{\rm Tr}\left( \hat{\tau}_{3}\otimes \hat{\sigma}%
_{0}\right) \int {\rm d}\epsilon \left( \check{g}^{R}.\partial _{x}\check{g}%
^{K}+\check{g}^{K}.\partial _{x}\check{g}^{A}\right) \,.  \label{current}
\end{equation}

In order to find the Green's functions $\check{g}^{R(A)}$, we multiply the
components (1,1) and (2,2) of the matrix equation (\ref{usadel}) (the Usadel
equations) by the density of states $\nu _{F,S}$ in the F and the S layers
respectively, and integrate over the thickness of the bilayers. Neglecting
the influence of one bilayer on the other (this means that $\left( \check{g}%
\partial _{x}\check{g}\right) =0$ at the F/I interface), we obtain, in the
Matsubara representation, the
following equation:

\begin{equation}
 \left[ \check{M},\check{g}\right] +\left[ \hat{%
\Delta}_{eff}\otimes \hat{\sigma}_{3},\check{ g}\right] =0\,\,,
\label{usa1}
\end{equation}
Here $\check{g}$ denotes the Matsubara Green's function, and the matrix
$\check{M}$ is given by
\begin{equation}
  \check{M}=\hat{\tau}_{3}\otimes (\hat{%
\sigma}_{0}i|\omega _{m}|+\hat{\sigma}_{3}h_{eff}{\rm sgn}\omega _{m}\cos \alpha )-%
\hat{\tau}_{0}\otimes \hat{\sigma}_{2}h_{eff}{\rm sgn}\omega _{m}\sin \alpha\; ,
\end{equation} 
where $\omega_m$ is the Matsubara frequency.
 In what follows we
 will skip the indices $eff$.
 We assume that the vector ${\bf h}$ in the left layer is
oriented along the z-axis and has the components $h\left( 0,\sin \alpha
,\cos \alpha \right) $ in the right electrode. One can simplify Eq. (\ref
{usa1}) in the right bilayer with the help of the following unitary transformation
\begin{equation}
\widetilde{\check{g}}=\check{U}^{+}.\check{g}.\check{U}\;,
\label{rotation}
\end{equation}
where $\check{U}=\hat{\tau}_{0}\otimes \hat{\sigma}_{0}\cos (\alpha
/2)+i\sin (\alpha /2)\hat{\tau}_{3}\otimes \hat{\sigma}_{1}$.
In this case one obtains for the both layers the same equation: 
\begin{equation}
\left[ \hat{\tau}_{3}\otimes (\epsilon \hat{\sigma}_{0}+h_{F}\hat{\sigma}%
_{3}),\check{g}\right] +\left[ \hat{\Delta}_{S}\otimes \hat{\sigma}_{3},%
\check{g}\right] =0\;.  \label{usadelrot1}
\end{equation}
We can solve Eq.(\ref{usadelrot1}) by making the ansatz 
\begin{equation}
\check{g}=\hat{\tau}_{3}\otimes (a_{0}\hat{\sigma}_{0}+a_{3}\hat{\sigma}%
_{3})+\hat{\Delta}_{S}\otimes (b_{0}\hat{\sigma}_{0}+b_{3}\hat{\sigma}%
_{3})\,.  \label{ansatz1}
\end{equation}
From Eq. (\ref{usadelrot1}) and the normalization condition (\ref
{normalization}) one can obtain the coefficients $a$'s and $b$'s. In the
left bilayer $\check{g}$ is given by the expression (\ref{ansatz1}) while in
the right bilayer it is given by $\check{g}^{(r)}=\check{U}^{+}\widetilde{%
\check{g}}^{(r)}\check{U}$, i.e. 
\[
\check{g}^{(r)}=\hat{\tau}_{3}\otimes (a_{0}\hat{\sigma}_{0}+a_{3}\cos
\theta \hat{\sigma}_{3})-\hat{\tau}_{0}\otimes a_{3}\sin \theta \hat{\sigma}%
_{2}+\hat{\Delta}_{S}\otimes (b_{0}\cos \theta \hat{\sigma}_{0}+b_{3}\hat{%
\sigma}_{3})-\hat{\tau}_{3}\hat{\Delta}_{S}\otimes ib_{0}\sin \theta \hat{%
\sigma}_{1}\,.
\]
According to Eq. (\ref{current}) only the coefficients $b_{0}$ and $b_{3}$
will enter in the expression for the Josephson current, and they are given
by 
\[
(b_{3})_{l,r}=\frac{1}{2}\left( \frac{1}{\xi _{+}}+\frac{1}{\xi _{-}}\right)
_{l,r}\;{\rm and}\;(b_{0})_{l,r}=\frac{1}{2}\left( \frac{1}{\xi _{+}}-\frac{1}{%
\xi _{-}}\right) _{l,r}\;,
\]
where $\xi _{\pm }=\sqrt{\epsilon _{\pm }^{2}-|\Delta _{S}|^{2}}$, and $%
\epsilon _{\pm }=i\omega _{m}\pm h$. By writing $\Delta _{S}=|\Delta
_{S}|\exp (i\varphi )$ in the right side one obtains the following
expression for the critical current 
\begin{equation}
eV_{c}(\alpha )\equiv eI_{c}R_{b}=2\pi T\Delta _{l}\Delta
_{r}\sum_{m>0}\left\{ {\rm Re}\left( \frac{1}{\xi _{m}}\right) _{l}{\rm Re}%
\left( \frac{1}{\xi _{m}}\right) _{r}-\cos \theta {\rm Im}\left( \frac{1}{%
\xi _{m}}\right) _{l}{\rm Im}\left( \frac{1}{\xi _{m}}\right) _{r}\right\} \;,
\label{crit_current_sfisf}
\end{equation}
where $\xi _{n}=\sqrt{\left( \omega _{m}+ih_{F}\right) ^{2}+\Delta _{S}^{2}}$
and $R_{b}$ is the tunnel resistance of the I layer. We note that the system
under consideration is equivalent to a Josephson junction consisting of two
magnetic superconductors. 
 The same structure was also analyzed in Ref.\cite{Koshina}, where the
critical current was calculated for different S/F interface transparencies.
The authors have found the conditions under which the system undergoes a
transition to the $\pi $ state; however they analyzed only the case of
parallel magnetization. 

Here we consider two limiting cases: a) a parallel relative orientation of
the magnetizations, i.e. $\alpha =0$ and b) an antiparallel orientation: $%
\alpha =\pi $.

In the case $\alpha =0$ according to Eq. (\ref{crit_current_sfisf}), the
critical current is given by the expression 
\begin{equation}
eV_{c\uparrow \uparrow }\equiv eI_{c\uparrow \uparrow }R_{b}=4\pi T\Delta
_{S}^{2}\sum_{m}\frac{\omega _{m}^{2}+\Delta _{S}^{2}-h_{F}^{2}}{\left(
\omega _{m}^{2}+\Delta _{S}^{2}-h_{F}^{2}\right) ^{2}+4\omega
_{m}^{2}h_{F}^{2}}\;.  \label{crit_curr_para}
\end{equation}
In writing Eq. (\ref{crit_curr_para}) we assumed that $h_{F}$ and $|\Delta
_{S}|$ are the same in both bilayers (symmetric structure). The dependence
of the critical current on the exchange field $h_{F}$ is shown in Fig.2. At $T=0$ the current $I_{c}$ is constant up to the value $%
h_{F}=\Delta _{0}$ where it drops to zero; $\Delta _{0}$ is the effective
energy gap $\Delta _{S}$ at zero temperature and zero exchange field. This
is a consequence of the fact that the order parameter $\Delta $ is also
constant. We do not consider here a possible transition to the LOFF phase
predicted by Larkin and Ovchinnikov (LO) \cite{LOv} and Fulde and Ferrell
(FF) \cite{Fulde} for the region $0.755\Delta _{S0}<h_{F}$. We argue that
since the homogeneous superconducting state in this region is a metastable
state, its realization is possible. Nevertheless our result is definitely
valid for the region of small $h_{F}$, and a possible transition to the LOFF
would manifest itself in a drop of the the critical current.

More interesting is the case when the relative orientation of the
magnetizations is antiparallel, i.e. $\alpha =\pi $. This case was considered in
Ref. \cite{BVE}. Then, the critical
current is given by the expression 
\begin{equation}
eV_{c\uparrow \downarrow }(\pi )\equiv eI_{c\uparrow \downarrow }R_{b}=4\pi
T\Delta _{S}^{2}\sum_{m}\frac{1}{\sqrt{\left( \omega _{m}^{2}+\Delta
_{S}^{2}-h_{F}^{2}\right) ^{2}+4\omega _{m}^{2}h_{F}^{2}}}\;.
\label{crit_curr_anti}
\end{equation}
In this case the dependence of $I_{c}$ on $h_{F}$ is completely different
from that given by Eq. (\ref{crit_curr_para}) (see Fig.3). The critical current
determined by Eq. (\ref{crit_curr_anti}) increases with increasing $h_{F}$
(i.e. with increasing either $h$ or $d_{F}$) and even diverges at zero
temperature when $h_{F}\rightarrow \Delta _{S}$. Of course, there is no real
divergence of $I_{c}$ since, for example, finite temperatures  smear out
this divergency. The critical current has a maximum at some value of $h_{F}
$ close to $\Delta _{0}$. With decreasing $T$ the maximum value of $I_{c}$
increases and its position is shifted towards $\Delta _{0}$. For arbitrary
relative orientations of magnetizations the expression for $V_{c}(\alpha )$
can be presented in the form 
\begin{equation}
V_{c}(\alpha )=V_{c\uparrow \uparrow }\cos ^{2}(\alpha /2)+V_{c\uparrow
\downarrow }\sin ^{2}(\alpha /2)\;.  \label{generalalpha}
\end{equation}
Therefore, the singular part is always present and its contribution reaches
100\% at $\alpha =\pi $.

All the conclusions given above remain valid also for two magnetic
superconductors with uniformly oriented magnetizations in each layer. As in the previous case of parallel
orientations, the state with $h_{F}=\Delta _{0}$ might be unreachable for
the antiparallel orientation due to the appearance of the inhomogeneous LOFF
state. However the singular behavior of $I_{c}$ can be realized at smaller
values of $h$ in a
structure with large enough S/F interface resistance $R_{S/F}$. In this case
the bulk properties of the S film are not changed by the proximity of the F
film (to be more precise the condition $R_{S/F}>(\nu_Fd_F/\nu_Sd_S)\rho_{F}\xi_{F}$ must be satisfied; $\rho_F$ is the specific resistance of the F film). Then, as one can readily show \cite{mcmillan}, a subgap $\epsilon
_{sg}=\left( D\rho \right) _{F}/\left( R_{S/F}d_{F}\right) $ arises in the F
layer. The Green's functions in the F layer have the same form as in Eq. (%
\ref{ansatz1}) with $\Delta _{S}$ replaced by $\epsilon _{sg}$. The
singularity in $I_{c}(h_{F})$ occurs at $h_{F}$ equals to $\epsilon _{sg}$,
and the LOFF state does not arise because the subgap $\epsilon _{sg}$ is not
determined by the self-consistency equation.

Another model of a SF/I/FS Josephson junction was considered in a recent work
\cite{koshina2}. The authors assumed that $d_s>\xi_s$ and that the conductance
of the S layer in the normal state is much larger than the conductance of the
F layer. According to this model the critical current may either increase or
decrease as a function of $h$, and for certain values of $h$ may become
negative ($\pi$-junction). It is worthwhile noting that real structures are
much more complicated than the models presented above. In order to obtain a
complete description of the systems one should take into account that
electrons with up and down spins have different density-of-states,
conductivities and transmission coefficient through the S/F interface.

\section{Triplet pairing and long-range proximity effect}

Although in almost all known superconductors Cooper pairs are in spin singlet
state, the triplet superconductivity have been studied in some works. Many years
ago Berezinskii studied a possible triplet phase in superfluid $^3He$
\cite{Berez}. The triplet component of the condensate function proposed
in Ref.\cite{Berez} was odd in frequency and even in momentum. However
nowadays it is known that this hypothetical condensate function does not take place
 in $^3He$. The discovery of superconductivity in $Sr_2RuO_4$
\cite{Maeno} has arisen the general interest in triplet superconductivity.  In
the case of S/F structures, the possible role of a triplet component in
transport  properties was studied in Refs. \cite{Falko,Zhou}. In both cases
the triplet component arose as a result of mesoscopic fluctuations, and the
correction to the conductance were very small. In this section we suggest
another mechanism of formation of triplet pairing in S/F structures. This is
due to a local inhomogeneity of the magnetization in the vicinity of the S/F
interface. 

We consider the system shown in Fig.4 and assume that the magnetization
orientation varies linearly from $\alpha =0$ at $x=0$ to $\alpha_{w}= Qw$ at 
$x=w$. Here $\alpha$ is the angle between $M$ and the $z$-axis (the $x$-axis is parallel to the f
wire). Thus in this region the magnetization is given by
\begin{equation}
  \label{eq:magnetization}
  {\bf M}=h\left(0, \sin Qx, \cos Qx\right)\; . 
\end{equation}
We also consider the diffusive limit corresponding to short mean free path
and to the condition $h\tau\ll 1$. Thus we may describe the system using
Eq. \ref{usadel}. This equation contains the normal $\check{g}$ and anomalous
$\check{f}$ Green's functions, which are $4\times 4$ matrices in the Nambu$%
\otimes $spin space. Due to the strong mismatch of the Fermi surfaces the
transmission coefficient through the S/F interface is small and therefore we
can assume that the anomalous condensate function $\check{f}$ is also
small. In the case of high transparency the order parameter in the
superconductor might be suppressed, and therefore it is also possible to assume
a weak proximity effect. Thus in both cases  we can use the linearized Usadel equation for the retarded
matrix (in spin space) Green's function $\hat{f}^R$, which has the form(the index $R$ is
dropped)

\begin{equation}
-iD\partial _{{\bf r}}^{2}\hat{f}+2\epsilon \hat{f}-2\Delta \hat{\sigma}%
_{3}+\left( \hat{f}\hat{V}^{\ast }+\hat{V}\hat{f}\right) =0 \; . \label{a2}
\end{equation}
Here the matrix $\hat{V}$ is defined as $%
\hat{V}=h\left( \hat{\sigma}_{3}\cos \alpha +\hat{\sigma}_{2}\sin \alpha
\right) $, where $\alpha $ varies with $x$ as shown in Fig.4.  Eq. (\ref{a2}) is supplemented by the boundary conditions at the interface
that can also be linearized \cite{Zaitsev}. Assuming that there are no spin-flip processes at
the S/F interface, we have 
\begin{equation}
\left. \partial _{x}\hat{f}\right| _{x=0}=\left( \rho /R_{b}\right) \hat{f}%
_{S}\;,  \label{bound-cond}
\end{equation}  
where $\rho $ is the resistivity of the ferromagnet, $R_{b}$ is the S/F
interface resistance per unit area in the normal state, and $f_{S}=\hat{%
\sigma}_{3}\Delta /\sqrt{\epsilon ^{2}-\Delta ^{2}}$. 

The solution of Eq. (\ref{a2}) is trivial in the superconductor but needs
some care in the ferromagnet. In the region $0<x<w$ the solution $\hat{f}$
can be sought in the form
\begin{equation}
\hat{f}=\hat{U}\left( x\right) \hat{f}_{n}\hat{U}
\left( x\right)\; ,\label{rotation2}
\end{equation}
 where $\hat{U}$ is again an unitary transformation given by $\hat{U}=\left( x\right) =\hat{\sigma}_{0}\cos \left(
Qx/2\right) +i\hat{\sigma}_{1}\sin \left( Qx/2\right)$.

 Substituting Eq. (\ref{rotation2}) into Eq. (\ref{a2}) and assuming that the solution depends on the
coordinate $x$ only we obtain the following equation for $\hat{f}_{n}$ 
\begin{equation}
-iD\partial_{xx}^2\hat{f}_n\!\!+\!\!i\left(DQ^2/2\right)\left(\hat{f}_n+\hat{%
\sigma}_1 \hat{f}_n\hat{\sigma}_1\right)\!\!+\!\!DQ\left\{\partial_{x}\hat{f}%
_n,\hat{\sigma}_1\right\}  
+2\epsilon\hat{f}_n+h\left\{\hat{\sigma}_3,\hat{f}_n\right\}\!=0\; .  \label{a5}
\end{equation}
Here $\{...\}$ is the anticommutator. In the region $x>w$,  $\hat{f}_{n}$
satisfies Eq. (\ref{a5}) with  $Q=0$.

We see from Eq. (\ref{a5}) that   the singlet and triplet components of the
anomalous function $\hat{f}_{n}$ inevitably coexist in the ferromagnet. They are
mixed by the rotating exchange field $h$. In the region $x>w$ these components decouple and their amplitudes should be found by
matching the solutions at $x=w$. 

Eq. (\ref{a5}) can be solved exactly. The solution $\hat{f}_{n}$ can be
written in the form 
\begin{equation}
\hat{f}_{n}= \hat{\sigma}_0A\left( x\right) +\hat{\sigma}_{3}B\left(
x\right)+i\hat{\sigma}_{1}C\left( x\right)  \label{2}
\end{equation}

The function $C\left( x\right) $ in Eq. (\ref{2}) is the amplitude of the
triplet pairing, whereas the first and the second term describe the singlet
one. Substituting Eq. (\ref{2}) into Eq. (\ref{a5}) we obtain a system of
three equations for the functions $A $, $B$
and $C $, which can be sought in the form 
\begin{equation}
A\left( x\right) =\sum_{i=1}^{3}\left( A_{i}\exp \left( -\kappa _{i}x\right)
+\bar{A}_{i}\exp \left( \kappa _{i}x\right) \right)  \label{3}
\end{equation}
The functions $B(x)$ and $C(x)$ can be written in a similar way. The
eigenvalues $\kappa _{i}$ obey the algebraic equations 

\begin{eqnarray}
\left( \kappa ^{2}-\kappa _{\epsilon }^{2}-Q^{2}\right) C-2\left( Q\kappa
\right) A &=&0  \nonumber \\
\left( \kappa ^{2}-\kappa _{\epsilon }^{2}\right) B-\kappa _{h}^{2}A &=&0
\label{4} \\
\left( \kappa ^{2}-\kappa _{\epsilon }^{2}-Q^{2}\right) A-\kappa
_{h}^{2}B+2\left( Q\kappa \right) C &=&0\;,  \nonumber
\end{eqnarray}
where $\kappa _{\epsilon }^{2}=-2i\epsilon /D$ and $\kappa _{h}^{2}=-2ih/D$
(indices $i$ were dropped).
The eigenvalues $\kappa $ are the values at which the determinant of Eqs. (%
\ref{4}) turns to zero. From the first equation of Eqs. (\ref{4}) we see
that in the homogeneous case ($Q=0$) the triplet component has a
characteristic penetration length $\sim \kappa _{\epsilon }^{-1}$, but we
see from Eq. (\ref{bound-cond}) that its amplitude is zero. If $Q\not=0$,
the triplet component $C$ is coupled to the singlet component ($A$, $B$)
induced in the ferromagnet according to the boundary condition Eq. (\ref
{bound-cond}) (proximity effect). If the width $w$ is small, the triplet
component changes only a little in the region $(0,w)$ and spreads over a
large distance of the order $\left| \kappa _{\epsilon }^{-1}\right| $ in the
region $(0,L)$. In the case of a strong exchange field $h$, $\xi _{F}$ is
very short ($\xi _{F}\ll w,\xi _{T}$), the singlet component decays very
fast over the length $\xi _{F}$, and its slowly varying part turns out to be
small. In this case the first two eigenvalues $\kappa _{1,2}\approx(1\pm
i)/\xi _{F}$ can be used everywhere in the ferromagnet $\left( 0<x<L\right) $%
, where $L$ is the length of the ferromagnet. As concerns the third
eigenvalues, we obtain $\kappa _{3}=\sqrt{\kappa _{\epsilon }^{2}+Q^{2}}$ in
the interval $(0,w)$, and $\kappa _{3}=\kappa _{\epsilon }$ in the interval $%
(w,L)$. The amplitude $B_{3}$ of the slowly varying part of the singlet
component is equal to $B_{3}=2\left( Q\kappa _{3}/\kappa _{h}^{2}\right)
C_{3}\ll C_{3}$.

All the amplitudes should be chosen to satisfy the boundary conditions at $%
x=0$ (Eq. (\ref{bound-cond})) and zero boundary condition at $x=L$.  For
the triplet component we obtain (we restore the indices R(A)) 
\begin{equation}
C^{R(A)}(x)=\mp i\left\{ QB(0)\sinh \left( \kappa _{\epsilon
}(L-x)\right)  \left[ \kappa _{\epsilon }\cosh \Theta _{\epsilon }\cosh
\Theta _{3}+\kappa _{3}\sinh \Theta _{\epsilon }\sinh \Theta _{3}\right]
^{-1}\!\right\} ^{R(A)}\; ,  \label{5}
\end{equation}
where $w<x<L$, $B^{R(A)}(0)\!\!\!=\!\!\left( \rho \xi _{h}/2R_{b}\right)
f_{S}^{R(A)}$ is the amplitude of the singlet component at the S/F
interface, $\Theta _{\epsilon }\!\!\!=\!\!\kappa _{\epsilon }L$, $\Theta
_{3}\!\!\!=\!\!\kappa _{3}w$, and $\kappa _{\epsilon }^{R(A)}\!\!\!=\!\!%
\sqrt{\mp 2i\epsilon /D}$.

It is clear from Eq. (\ref{5}), that the triplet component is of the same
order of magnitude as the singlet one at the interface. Indeed, for the case 
$w\ll L$ we obtain from Eq. (\ref{5}) $\left| C(0)\right| \sim B(0)/\sinh
\alpha _{w}$, where $\alpha _{w}=Qw$ is the angle characterizing the
rotation of the magnetization. Therefore if the angle $\alpha _{w}\leq 1$
and the S/F interface transparency is not too small, the singlet and triplet
components are not small. They are of the same order in the vicinity of the
S/F interface, but while the singlet component decays abruptly over a short
distance ($\sim \xi _{F}$), the triplet one varies smoothly along the
ferromagnet, turning to zero at the F reservoir. In Fig.5 we plot the
spatial dependence of the singlet $|B(x)|$ and the triplet $|C(x)|$
components for two different $Q$. One can see that the singlet component
decays abruptly undergoing the well known oscillations \cite{Buzdin2} while
the triplet one decays to zero slowly. This decay in the region $(0,w)$
increases with increasing $Q$.

Thus, we come to a remarkable conclusion: the penetration of the
superconducting condensate into a ferromagnet may be similar to the
penetration into a normal metal. The only difference is that, instead of the
singlet component in the case of the normal metal, the triplet one
penetrates into the ferromagnet. Of course, in order to induce the triplet
component one needs an inhomogeneity of the exchange field at the interface.

The presence of the condensate function (triplet component) in the
ferromagnet can lead to interesting long-range effects. One of them is a
change of the conductance of a ferromagnetic wire in a S/F structure (see
inset in Fig.4) when the temperature is lowered below $T_{c}$. This effect
was observed first in S/N structures and later was successfully explained
(see, e.g. reviews \cite{Been,Lamb}). Now we consider the S/F structure
shown in the inset of Fig.1. The normalized conductance variation $\delta 
\tilde{G}=\left( G-G_{n}\right) /G_{n}$ is given by the expression \cite{VZK}%
: 
\begin{equation}
\delta \tilde{G}=-\frac{1}{32T}{\rm Tr}\int {\rm d}\epsilon F_{V}^{\prime
}\left\langle \left[ \hat{f}^R(x)-\hat{f}^A(x)\right] ^{2}\right\rangle \;.
\label{6}
\end{equation}
Here $G_{n}$ is the  conductance in the normal state, $<..>$ denotes the average over the length of the ferromagnetic wire
between the F reservoirs, and $F_{V}^{\prime }$ is given by the expression 
\begin{equation}
  F_{V}^{\prime }=1/2
\left[ \cosh ^{-2}((\epsilon +eV)/2T)+\cosh^{-2}((\epsilon -eV)/2T)\right]\; .
\end{equation} 
The function $\hat{f}$ is given by the third term
of Eq. (\ref{2}) with $C^R=-\left(C^A\right)^*$ (we neglect the small
singlet component). Substituting Eqs. (\ref{2}, \ref{5}) into Eq. (\ref{6})
one can determine the temperature dependence $\delta \tilde{G}\left(
T\right) $. Fig.6 shows this dependence. We see that $\delta \tilde{G}$
increases with decreasing temperature and saturates at $T=0$. This monotonic
behavior of $\delta \tilde{G}$ contrasts with the so called reentrant
behavior of $\delta \tilde{G}$ in S/N structures \cite{Art,Nazarov} and is
a result of broken time-reversal symmetry of the system under consideration.

Available experimental data are still controversial. It has been established
in a recent experiment \cite{Chandr} that the conductance of the
ferromagnet does not change below $T_{c}$ and all changes in $\delta G$ are
due to changes of the S/F interface resistance $R_b$. However, in other
experiments $R_b$ was negligibly small \cite{Petr}. The mechanism
suggested in our work may explain the long-range effects observed in the
experiments \cite{Petr,Pann}. At the same time, the result of the
experiment \cite{Chandr} is not necessarily at odds with our findings. The
inhomogeneity of the magnetic moment at the interface, which is the crucial
ingredient of our theory, is not a phenomenon under control in these
experiments. One can easily imagine that such inhomogeneity existed in the
structures studied in Refs. \cite{Petr,Pann} but was absent in
those of Ref. \cite{Chandr}. The magnetic inhomogeneity near the interface
may have different origins. Anyway, a more careful study of the possibility
of a rotating magnetic moment should be performed to clarify this question.

In order to explain the reentrant behavior of $\delta G(T)$ observed in
Refs. \cite{Petr,Pann} one should take into account other
mechanisms, as those analyzed in Refs. \cite{Naz,Falko,Golubov}. However,
this question is beyond the scope of the present paper.

We note that at the energies $\epsilon $ of the order of Thouless energy $%
\epsilon \sim E_{T}$ the triplet component spreads over the full length $L$
of the ferromagnetic wire (see Fig.2). This long-range effect differs
completely from the proximity effect in a ferromagnet with a uniform
magnetization considered recently in Ref.\cite{Buzdin}. In the latter case
the characteristic wave vector is equal to $\kappa _{1,2}=\sqrt{-2i(\epsilon
\pm h)/D}$ (cf. Eqs. (\ref{4})). It was noted in Ref. \cite{Buzdin} that if $%
\epsilon\rightarrow\pm h$, then $\kappa_{1,2}\rightarrow 0$ and the singlet
component penetrates in the ferromagnet. If the characteristic energies $%
\epsilon_{ch}\sim E_T,T$ are much less than $h$, the penetration length $%
\left| \kappa _{1,2}\right| ^{-1}$ is of the order $\xi _{F}$ and is much
shorter then $\xi _{T}$ or $L$.

\section{Conclusion}

We analyzed two effects related to a nonhomogeneous magnetization in S/F
structures. First we have calculated the Josephson critical current  $I_{c}$
in a tunnel S/F-I-S/F junction. We have shown that in contrast to a common
view, in the case of an antiparallel configuration (the magnetization
vectors in the S/F electrodes  are antiparallel to each other) the critical
current $I_{c}$ increases with the increasing  exchange field $h$. This means
that in a S/F-I-F/S junction the current $I_{c}$ can be even greater than
that in a S-I-S junction with the same S electrodes. Secondly we have calculated the conductance of a
mesoscopic F wire attached to a superconductor S. We have shown that in the
presence of a local inhomogeneity near the S/F interface, both singlet and
triplet components of the condensate are created in the ferromagnetic wire
due to the proximity effect. The singlet component penetrates into the
ferromagnet over a short length $\xi _{F}$, whereas the triplet component
can spread over the full mesoscopic length of the ferromagnet. This
long-range penetration of the triplet component should lead to a significant
variation of the ferromagnet conductance below $T_{c}$.

\section*{Acknowledgment}

We thank SFB 491 {\it Magnetische Heterostrukturen} for financial support.

\clearpage

FIG.1: The S/F-I-F/S system.

\bigskip
\bigskip
%%%%%%%%%%%%%%%%%%%%%%%%%%%%%%%

%%%%%%%%%%%%%%%%%%%%%%%%%%%%
FIG.2:Dependence of the normalized critical current on $h$ for different
  temperatures in the case of a  parallel orientation. Here $eV_{c}=eRI_{c}$, $h_{F}$ is the
  effective exchange field, $t=T/\Delta_{0}$ and $\Delta_{0}$ is the
  superconducting order parameter at $T=0$ and $h=0$.
%%%%%%%%%%%%%%%%%%%%%%%%%%%%%%%%
\bigskip
\bigskip

%%%%%%%%%%%%%%%%%%%%%

FIG.3: Dependence of the normalized critical current on $h$ for different
  temperatures in the case of an  antiparallel orientation. Here $eV_{c}=eRI_{c}$, $h_{F}$ is the
  effective exchange field, $t=T/\Delta_{0}$ and $\Delta_{0}$ is the
  superconducting order parameter at $T=0$ and $h=0$.
%%%%%%%%%%%%%%%%%%%%%%%%%%%%

\bigskip
\bigskip

%%%%%%%%%%%%%%%%%%%%%%%%%%%%%%%%%%%%%%%%%%%

 FIG.4: Schematic view of the structure under consideration. In the inset is
 shown the structure, for which we calculate the conductance variation: two
 ferromagnetic wires connected to two ferromagnetic and two superconducting reservoirs.

%%%%%%%%%%%%%%%%%%%%%%%%%%%%%%%%%%%%%%%%%%%%%
\bigskip
\bigskip

FIG.5: Spatial dependence  of the singlet (dashed line) and triplet (solid line)
  components of $\left|\hat{f}\right|$ in the F wire for different values
  of $\alpha_w$. Here $w=L/5$, $\epsilon=E_T$ and $h/E_T=400$. $E_T=D/L^2$ is
  the Thouless energy.

\bigskip
\bigskip

FIG.6: The $\protect\delta G(T)$ dependence. Here $\protect\gamma =\protect%
\rho \protect\xi _{h}/R_{b}$. $\Delta /E _{T}\gg 1$ and $%
w/L=0.05$.

%%%%%%%%%%%%%%%%%%%%%%%%%%%%%%%%%%%%%

\clearpage

\clearpage
%%%%%%%%%%%%%%%%%%%%%%%%%%%%%%

\begin{figure}
\epsfysize = 7cm
\vspace{0.2cm}
\centerline{\epsfbox{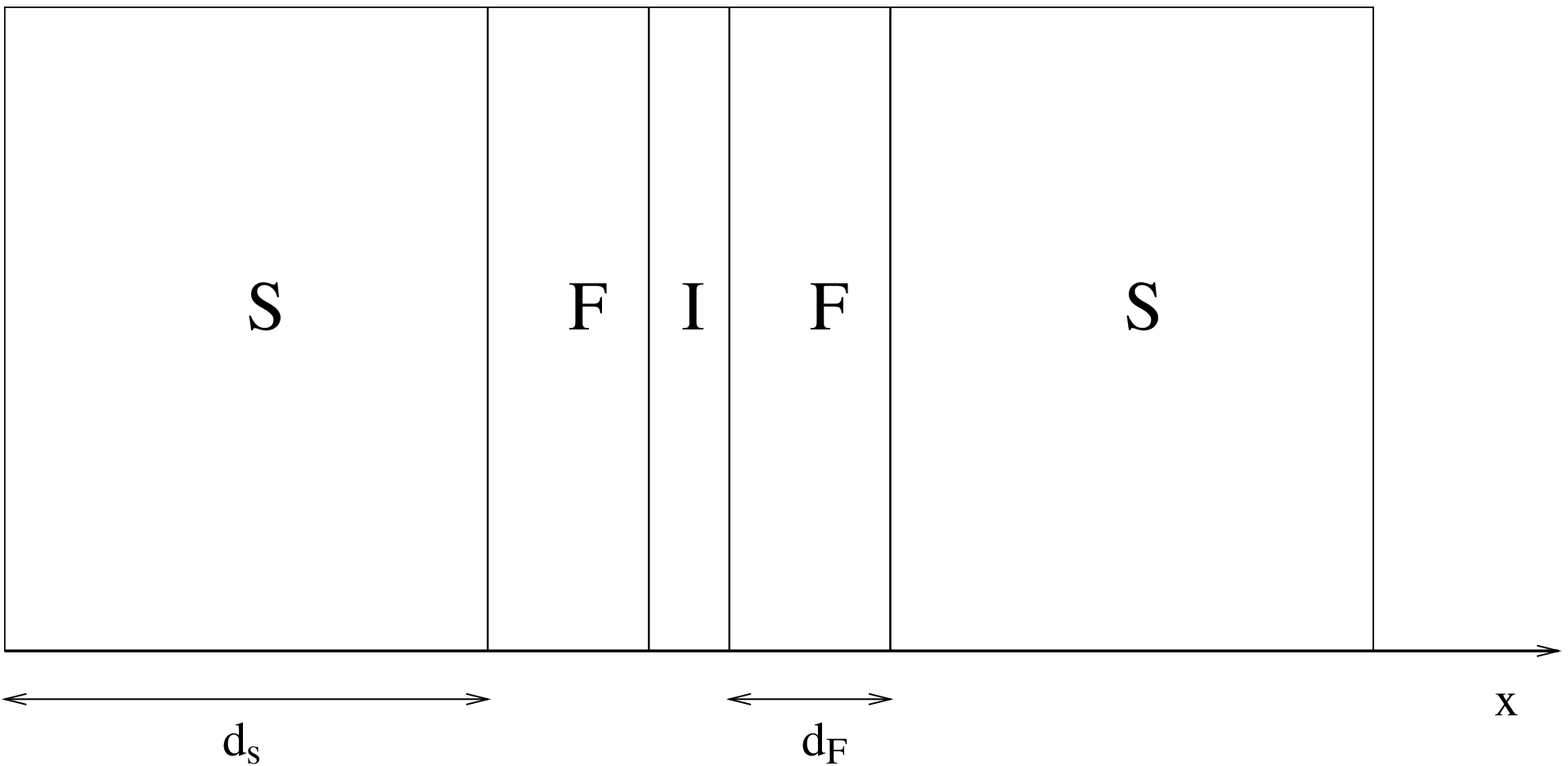
 }}
\vspace{2cm}
\caption{ }
\end{figure}

%%%%%%%%%%%%%%%%%%%%%%%%%%%%%%%
\clearpage
%%%%%%%%%%%%%%%%%%%%%%%%%%%%
\begin{figure}
\epsfysize = 16cm
\vspace{0.2cm}
\centerline{\epsfbox{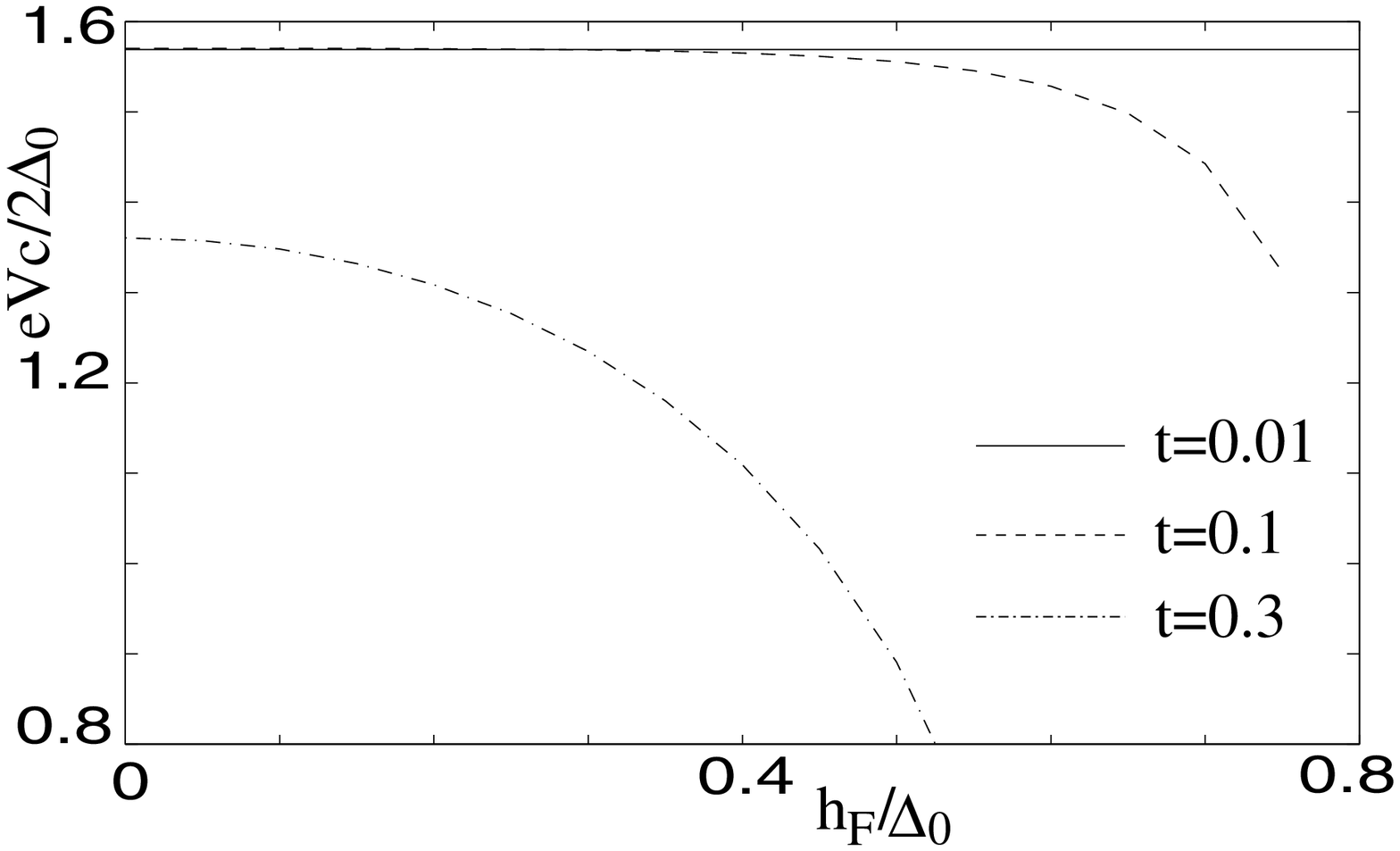
 }}
\vspace{0.2cm}
\caption{ }
\end{figure}
%%%%%%%%%%%%%%%%%%%%%%%%%%%%%%%%
\clearpage

%%%%%%%%%%%%%%%%%%%%%
\begin{figure}
\epsfysize = 16cm
\vspace{0.2cm}
\centerline{\epsfbox{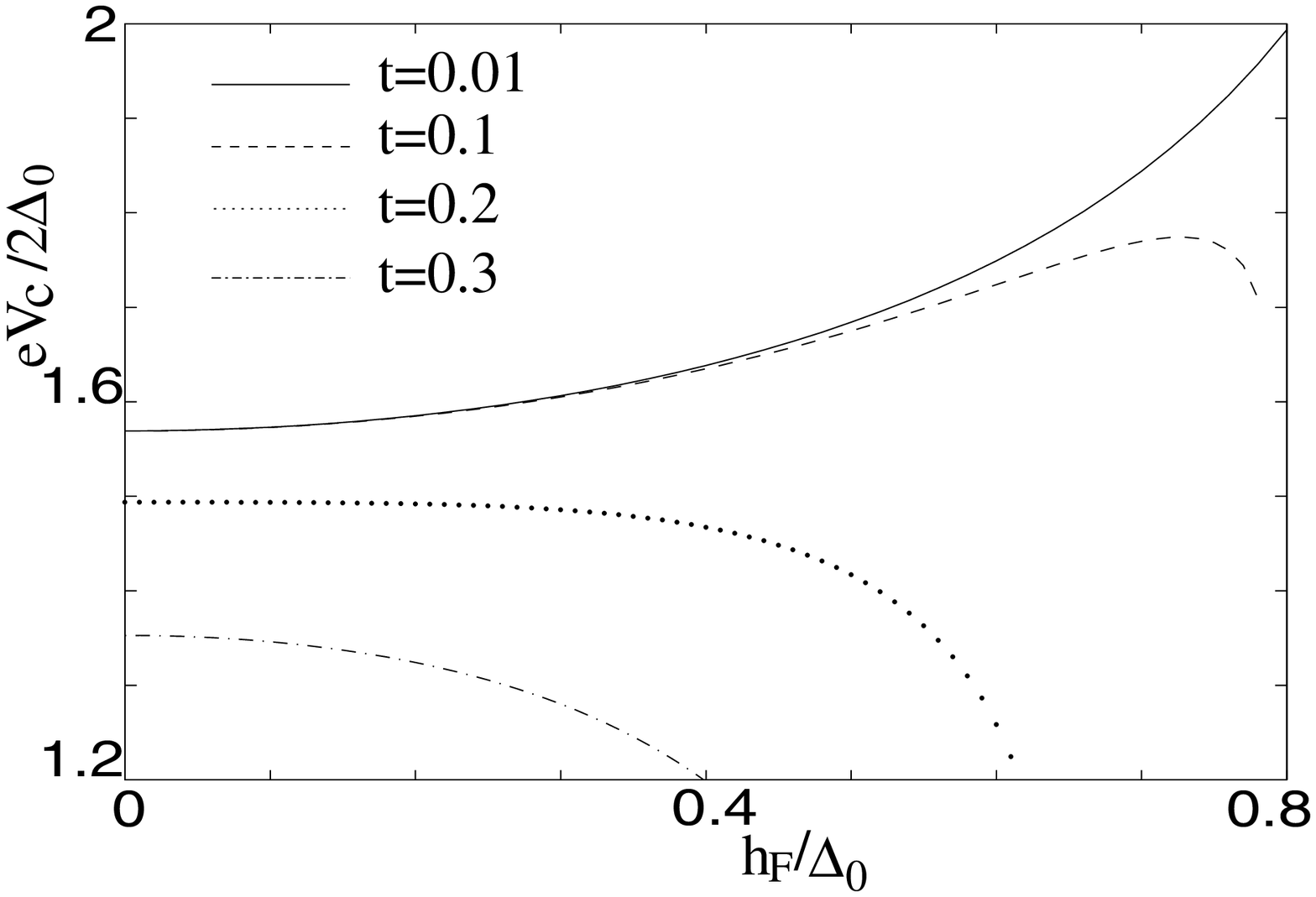
 }}
\vspace{0.2cm}
\caption{ }
\end{figure}
%%%%%%%%%%%%%%%%%%%%%%%%%%%%
\clearpage
%%%%%%%%%%%%%%%%%%%%%%%%%%%%%%%%%%%%%%%%%%%
\begin{figure}
\epsfysize = 10cm

\centerline{\epsfbox{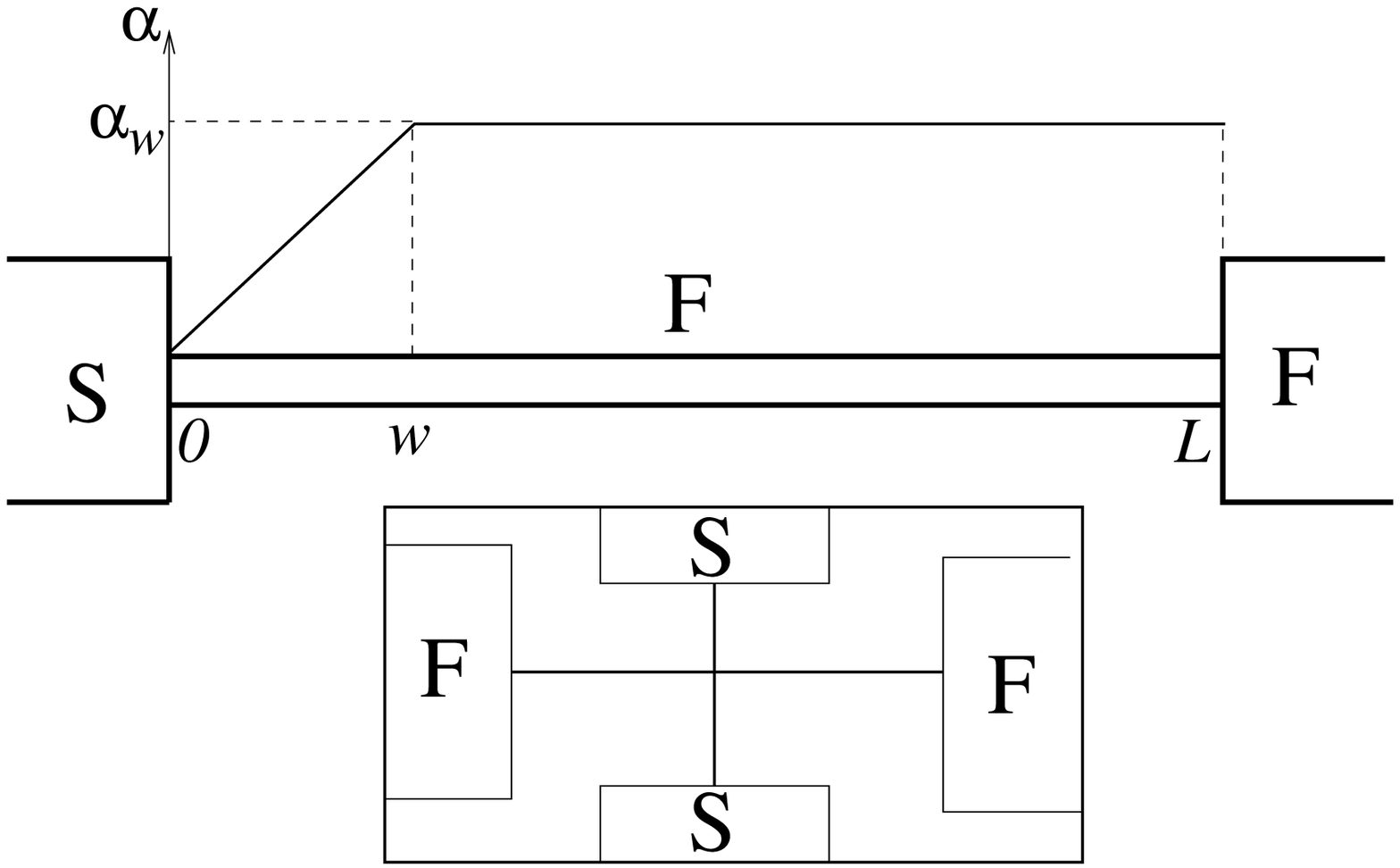
 }}
\vspace{0.2cm}
\caption{ }
\end{figure}

%%%%%%%%%%%%%%%%%%%%%%%%%%%%%%%%%%%%%%%%%%%%%
\clearpage

%%%%%%%%%%%%%%%%%%%%%%%%%%%%%%%%%
\begin{figure}
\epsfysize = 10.4cm

\centerline{\epsfbox{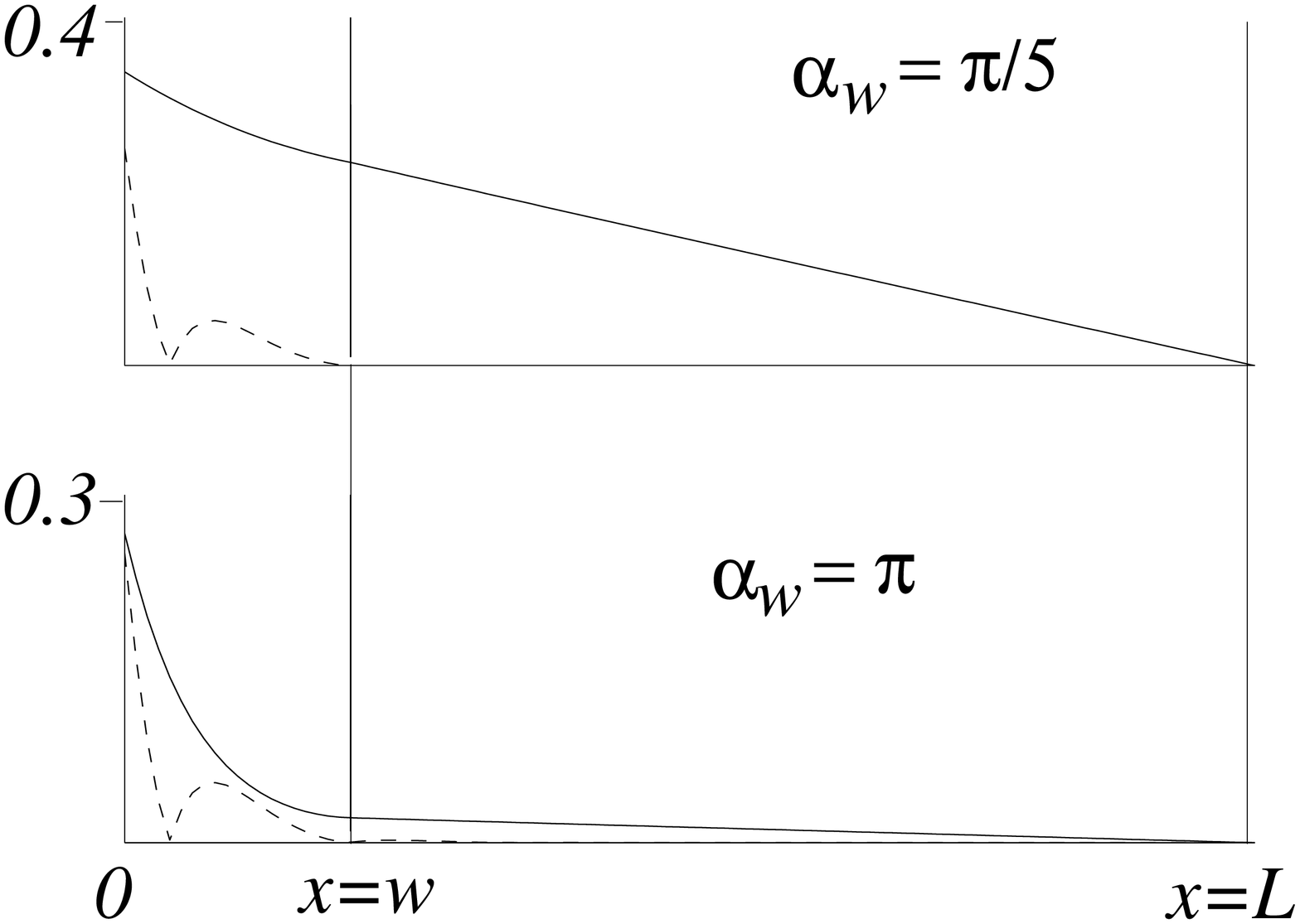
 }}
\vspace{0.2cm}
\caption{ }
\end{figure}

%%%%%%%%%%%%%%%%%%%%%%%%%%%%%%%%%
\clearpage

%%%%%%%%%%%%%%%%%%%%%%%%%%%%%%
\begin{figure}
\epsfysize = 16cm
\centerline{\epsfbox{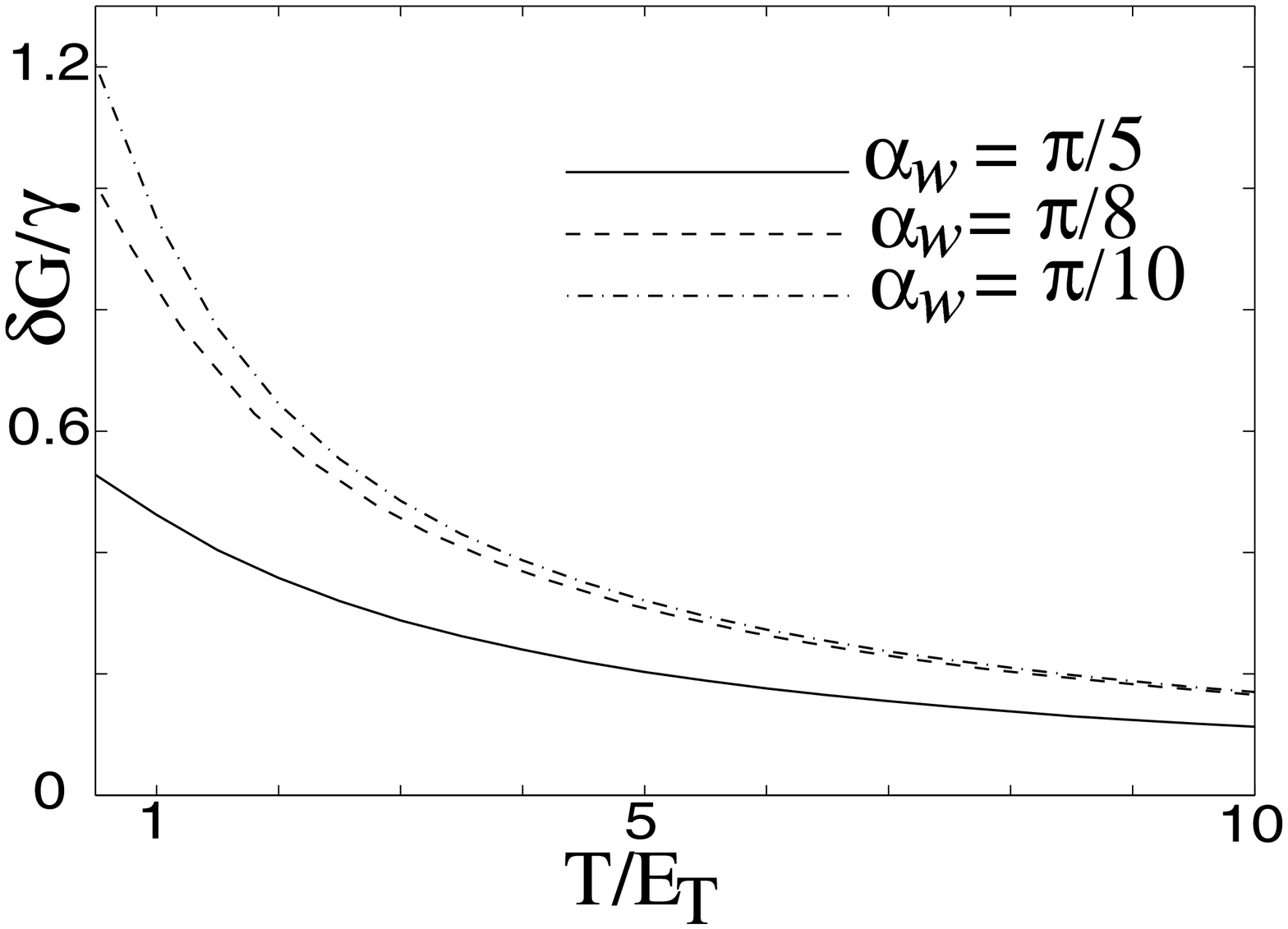
 }}
\vspace{0.2cm}
\caption{ }
\end{figure}
%%%%%%%%%%%%%%%%%%%%%%%%%%%%%%%%%%%%%

\end{document}